# Proton Conducting Graphene Oxide Coupled Neuron Transistors for Brain-Inspired Cognitive Systems

*Changjin Wan [1, 2], Liqiang Zhu [2], Yanghui Liu [2], Ping Feng [1], Zhaoping Liu [2], Hailiang Cao [2], Peng Xiao [2], Yi Shi [1*], and Qing Wan,[1, 2, *]*.

[1] School of Electronic Science & Engineering, and Collaborative Innovation Center of Advanced Microstructures, Nanjing University, Nanjing 210093, China

[2] Ningbo Institute of Material Technology and Engineering, Chinese Academy of Sciences, Ningbo 315201, China

**Abstract|** Neuron is the most important building block in our brain, and information processing in individual neuron involves the transformation of input synaptic spike trains into an appropriate output spike train. Hardware implementation of neuron by individual ionic/electronic hybrid device is of great significance for enhancing our understanding of the brain and solving sensory processing and complex recognition tasks. Here, we provide a proof-of-principle artificial neuron based on a proton conducting graphene oxide (GO) coupled oxide-based electric-double-layer (EDL) transistor with multiple driving inputs and one modulatory input terminal. Paired-pulse facilitation, dendritic integration and orientation tuning were successfully emulated. Additionally, neuronal gain control (arithmetic) in the scheme of rate coding is also experimentally demonstrated. Our results provide a new-concept approach for building brain-inspired cognitive systems.

[*] E-mail: wanqing@nju.edu.cn; yshi@nju.edu.cn

Our brain is a highly parallel, energy efficient and event-driven information processing system, which is fundamentally different from traditional von Neumann computers[1-3]. At single neuron level, information processing involves the transformation of input synaptic spike trains into an appropriate output spike train. In the beginning, individual neuron was regarded as the simple linear summation and thresholding device, and even low level computations, such as multiplication of signals, had to be carried out by groups of neurons[4, 5]. However, mounting theoretical work suggested that single neuron could act as more powerful computational unit[5-7]. Neurons have elaborate dendritic trees that receive thousands of synaptic inputs, and dendritic signal integration is one of the fundamental building blocks of information processing in the brain. Dendritic processing is highly nonlinear, and such dendritic nonlinearities enhance processing capabilities at the single-neuron level[8, 9].

The idea of building bioinspired solid-state devices has been around for decades. In the year of 1992, Shibata et al. proposed a new-concept neuron MOS transistor with multiple input gates, and the summation of gate voltage signals could be carried out by the charge sharing among multiple gate capacitors[10]. Up to now, such Si-based neuron transistors coupled by traditional gate dielectrics were mainly used for digital logic circuits and chemical sensors, and spiking neuromorphic applications were not investigated. In the past five years, artificial synapses have been demonstrated by a broad spectrum of two terminal devices, such as $Ag_2S$ atom switch, $WO_x$ memristor, $HfO_x$-based resistive switching, and $Ge_2Sb_2Te_5$-based phase change switch[11-15]. More recently, artificial synapses based on three-terminal transistors gated by ion-correlated materials such as ionic liquid and solid electrolyte films have also attracted considerable attention[16-20]. Ion conducting electrolyte films are favorable for electric-double-layer (EDL) modulation and enable the EDL transistors to emulate the synaptic plasticity and computing functions with very low energy consumption. Artificial neuron devices with multiple presynaptic inputs are of great significance for neuromorphic systems because biological neurons usually receive thousands of synaptic inputs and dendritic integration can greatly enrich the computational power of neurons. Here we provide, for the first time, proof-of-principle artificial neurons with multiple presynaptic inputs based on proton conducting graphene oxide (GO) electrolyte films coupled oxide-based EDL transistors. Such artificial neuron devices show dendritic integration, orientation tuning and neuronal gain control functions, which may provide a new-concept approach for brain-inspired cognitive systems.

GO is the derivative of a graphene sheet that has been modified with oxide functional groups[21, 22]. The epoxy, hydroxyl, and carboxyl groups are widely accepted as the main functional groups on both its 2D basal planes and edges, which endow GO with various technological applications such as supercapacitor, ion-exchange membrane, etc[23-25]. Figure 1a shows the X-ray photoelectron spectroscopy (XPS) C1s spectra of the as-prepared GO film. The C1s spectra were compared by deconvoluting each spectrum into four peaks that correspond to the following functional groups: carbon $sp^2$ (C=C, 284.8 eV), epoxy/hydroxyls (C-O, 286.2 eV), carbonyl (C=O, 287.8 eV), and carboxylates (O-C=O, 289.0 eV)[26, 27]. All percentages of the oxidized materials were combined such that GO had 66% oxidized carbon and 34% graphitic carbon indicating the abundant oxide functional groups. The proton conducting properties of GO-based materials were reported recently[28]. The protolysis process was proposed occurring through hydrogen bonding networks constructed by the oxygen containing functional groups and water molecules attached on the surface through a Grotthuss mechanism[28]. Figure 1b shows the possible mechanisms of propagation of protons supported by single and double GO walls[28, 29]. What's more, for multilayer GO, protons can change path from one layer to the surrounding layers through nanopores, which also contribute to the good proton conductivity[29]. Under the external electric field, an oriented transport of protons through a sequence of hopping results in the accumulation of protons at the GO/Au interface (bottom ITO positively biased), and a so-called EDL capacitor can be formed as the equilibrium state. Frequency-dependent specific capacitance curve of the GO film was shown in Fig. 1c. The specific capacitance was measured with an Au/GO/ITO sandwich structure as shown in the inset of Fig. 1c. A maximum specific capacitance of 18 $\mu F/cm^2$ is obtained at 1.0 Hz due to the EDL effect although the thickness of the GO is estimated to 540 nm. Figure 1d shows the leakage current of GO film, which is also measured with an Au/GO/ITO sandwich structure. The leakage current is estimated to be less than ~15 nA in the voltage range from -1.0 to 1.0 V. It's reported that GO has randomly allocated nonconductive $sp^3$ carbon sites that are responsible for abolishing the electrical conductivity of graphene[30]. So, proton conducting GO film can act as a good gate dielectric film for EDL modulation and new-concept device applications.

Figure 2a shows a schematic image of an artificial neuron with branched dendrites, where $w_i$ (i=1,2…n and m) denotes the synaptic weight of each input. Dendrites in a neuron can collect, integrate and modulate the thousands of presynaptic inputs and transmit the output spikes to other neurons through the axon[31]. Figure 2b schematically shows an indium-zinc-oxide (IZO) transistor Gated by GO film on ITO glass substrate with three in-plane gate electrodes ($G_1$, $G_2$ and $G_m$). Figure 2c shows the output characteristics ($I_{DS}$ vs $V_{DS}$) of the IZO-based neuron transistor. At low $V_{DS}$, $I_{DS}$ increases linearly with $V_{DS}$, indicating that the device has a good ohmic contact. Figure 2d shows the transfer characteristics of the transistors in the saturation region ($V_{DS}$=1.5 V . The gate voltage applied on the in-plane gate electrodes can be capacitively coupled to the IZO channel layer through two EDL capacitors ($C_1$ and $C_2$ in Fig. 2b) in series. Such in-plane gate neuron transistor exhibits a good electrical performance with a large current on/off ratio of $1.9 \times 10^6$ and a very small subthreshold swing of 84 mV/decade. A threshold voltage ($V_{TH}$) of ~0.5 V is estimated from the x-axis intercept of the square root of $I_{DS}$-versus-$V_{GS}$ plot, indicating the enhancement mode operation. A clear anticlockwise hysteresis window of ~0.2 V is observed due to the existence of mobile proton in the GO electrolyte film. The field-effect mobility ($\mu$) in the saturation region ($V_{DS}$>$V_{GS}$-$V_{TH}$) is estimated to be ~34 cm$^2$/V·s. The EDL can induce an extremely large electric-field on the order of megavolts per centimeter and result in a high density of charge carriers in the IZO channel. Such proton-related electrostatic modulation process is meaningful for low-voltage operation of EDL transistors and artificial neurons[18].

Excitatory postsynaptic currents (EPSCs) or inhibitory postsynaptic currents (IPSCs) can be triggered by the voltage spikes from the presynaptic neurons, which are collective processed to establish spatial and temporal correlated functions by the postsynaptic neuron[31]. The EPSC amplitude, i. e. the difference between the peak current and the base current, is used as the measure of the output for our artificial neurons. To mimic the EPSC, the presynaptic spikes were applied on the in-plane gate electrodes. The induced channel current is defined as the EPSC and is measured at $V_{DS}$=0.1 V. Figure 3a shows a typical EPSC triggered by a presynaptic spike (0.5 V, 10 ms). A sharp increase followed by a gradually decay of the current can be observed. The EPSC amplitude is estimated to be ~110 nA. A reasonable explanation was provided based on the EDL modulation. During a positive presynaptic spike, protons are electrically triggered and accumulated at the IZO/GO interface successively. The transient accumulation of protons then electrostatically induces a sharply increase of the channel current by the strong EDL coupling. After the spike, protons will diffuse back to their equilibrium positions due to the concentration gradient. The decay of proton concentration at the IZO/GO interface induces a gradually decay of channel current in turn. Paired-pulse facilitation (PPF) is an essential short-term plasticity in which an increased EPSC can be observed when the second spike closely follows a previous spike[32, 33]. The PPF can also be emulated in the artificial neuron. The inset of Fig. 3b shows the EPSCs in response to two successive presynaptic spikes (0.5 V, 10 ms) with a time interval of 40 ms. Two EPSC peaks ($A_1$ and $A_2$, respectively) were observed, and the second peak was obviously larger than the first one, indicating the PPF behavior. When the second pulse comes, the residual protons near the IZO channel triggered by the first spike will be added to the total proton concentration, which would induce an increment in EPSC as a result. Figure 3b systematically illustrates that the facilitation ratios ($A_2/A_1$) decrease with increasing of time interval ($\Delta T$). The facilitation ratios decrease gradually from 2.0 to 1.03 when the time interval increases from 20 ms to 200 ms.

A fundamental computation of neurons is the transformation of incoming synaptic information into specific patterns of synaptic output[34, 35]. An important step of this transformation is dendritic integration, which includes addition of unitary events occurring simultaneously in separate regions of the dendrite arbor (spatial summation) and addition of nonsimultaneous unitary events (temporal summation)[34]. Figure 3c schematically shows the spatial summation with two spatial isolated synaptic inputs. The red curves and blue curves are the EPSCs ($A_1$, $A_2$) triggered by single presynaptic spike ($V_1$ and $V_2$), respectively. The expected sum ($S_E$, green dashed curves) is defined as the arithmetic sum of two individual EPSC amplitudes ($A_1+A_2$). The measured sum ($S_M$, black curves) is the EPSC stimulated by the two simultaneously triggered spikes. Nonlinearity of spatial summation is called sublinear when the response to two or more inputs is always less than the sum of the individual responses ($S_M<S_E$), and called superlinear when the combined response always exceeds the linear prediction ($S_M>S_E$)[7]. The two presynaptic spikes ($V_1$ and $V_2$) ranged from 0.2 to 1.4 V were applied on $G_1$ and $G_2$, respectively. The inputs were first triggered individually and then triggered simultaneously, and thus the EPSC amplitudes ($A_1$, $A_2$ and $S_M$, respectively) can be measured successively. The measured sums were plotted as a function of the expected sum as plotted in Fig. 3d. Such spatial summation is nearly linear for low spike voltages, strong superlinear for intermediate spike voltages and sublinear for high spike voltages. For example, when $V_1=V_2=0.2$ V, $S_E=61$ nA and $S_M=78$ nA, when $V_1=V_2=1.0$ V, $S_E=5.6$ μA and $S_M=9.0$ μA, and when $V_1=V_2=1.4$ V, $S_E=12.6$ μA and $S_M=10.6$ μA. Such results are very similar to some biological experiments[7].

Neurons in the primary visual cortex respond preferentially to edges with a particular orientation[36, 37]. Such orientation selectivity is necessary for encoding of visual images orientations. The feed-forward filtering function is considered as the mechanism that cause the enhancement near the preferred orientation[37]. Since a higher PPF facilitation ratio is obtained with the shorter ΔT in our artificial neuron, orientation tuning function can also be realized in our artificial neuron. We built a simple model of visual system which consists of a photodetector and processing circuit connected to the gate electrode of the neuron transistor, as shown in Fig. 4a. The coordinate of the photodetector is (x, 0, 0). A square panel with five pairs of black-white grating patterns was moved along y axis in yoz plane. The orientation angle (θ) is defined as the angle between z axis and the grating orientation. For each orientation, the panel is moved from side to side at a speed of *v*, and the coordinate of the panel center is (0, y, 0). Each time when the edge of the grating pattern move across the coordinate origin is detected by the photodetector, the processing circuit would provide a voltage pulse (0.5 V, 10 ms) to the presynaptic terminal of the neuron transistor. The presynaptic spikes corresponding to each orientation were computed by Matlab software and generated by Keithley 4200. The detailed descriptions about the experiments were illustrated in **S1** and **S2** in supplementary information. As shown in Fig. 4b, when the orientation is $0^o$, ten presynaptic spikes with frequency of 50 Hz are triggered successively and induces a dramatically increased EPSC with the maximum value of ~319 nA. As shown in Fig. 4c, when the orientation angle is $78.5^o$, only two presynaptic spikes with a frequency of 10 Hz can be triggered and induce two EPSC peaks with a maximum EPSC amplitude of ~122 nA. The visual response is defined as the maximum EPSC amplitude at each orientation angle. The normalized visual responses for GO gated neuron transistor were plotted as a function of orientation angle as shown in Figure 4d, respectively. The experimental data were fitted by Gaussian function[38]. Followed a neuroscience protocol, the half-width ($W_H$) of the fitting curve measured at half-height were used as criterion of orientation selectivity[39]. The width is estimated to $44^o$ in this work, which is comparable to some biological experiments[39]. Although our model is not one-to-one correlation with biological visual systems, the demonstration of the orientation tuning in the proposed artificial neuron devices could provide important implications in capturing such neural computations by neuromorphic elements.

At last, the implementation of neuronal arithmetic in individual neuron transistor is discussed. A neuron with branched dendrites can collect, integrate and modulate the presynaptic inputs and transmit the output spikes to other neurons through the axon[31]. Two types of presynaptic inputs are correlated to such neural transmission: the driving inputs that can make the relevant neuron fire strongly, and the modulatory input that can tune the effectiveness of the driving input[40, 41]. As an analogy, the pulse voltages (0.5 V, 10 ms) were applied on $G_1$ and $G_2$ as driving inputs, and a low bias voltage (e. g. -0.1 and 0.1 V) was applied on $G_m$ as modulatory input. The EPSC can be triggered by the combined effects of these inputs, thus the EPSC amplitude is defined as neuronal output. Neuronal arithmetic refers to the multiplicative or additive modulation on input-output (I-O) relationship by tuning the modulatory input[5, 42, 43]. Figure 5a schematically shows the multiplicative and additive operations when the modulatory input is independent of the I–O relationship of the driving input. The algebraic transformation on neural I-O relationship is defined as multiplicative operation when the neuronal response can be written as a function of driving inputs multiplied by a function of modulatory inputs: R(d, m)=f(d)×g(m), where d and m are the driving input parameters and modulatory input parameters, respectively. The multiplicative operation can produce a change in the slope of the I–O relationship, which is also referred to a change in neuronal gain[42, 43]. Similarly, the neural response of additive operation is written as R(d,m)=f(d)+g(m).

The ways in which information is encoded are important for understanding the neuronal I–O relationship. The neuronal coding is generally classified into two categories: temporally correlated coding (correlation in the input timing) and rate coding (correlation in the input rate)[44, 45]. Figure 5b schematically shows the rate coding scheme in which information is encoded by the rate (*f*) of the driving inputs. In this work, the driving input parameter is defined by rate of the driving input and the modulatory input parameter is defined by the modulatory voltage, i. e. d=*f* and m=$V_m$. Figure 5c show the typical neural output in response to a Poisson distribution spike train as driving inputs (0.5 V, 10 ms) with rate of 40 spikes/s and a modulatory input of 0 V. The protocol of the Poisson distributed spike train is detailed illustrated in **S3** of the supplementary information. The neuronal output parameter is defined by EPSC amplitude and is estimated to 91 nA in this case. Figure 5d systematically shows the neural I-O relationship tuned by modulatory input voltages of -0.1, 0 and 0.1 V, respectively. For each case, the rates are varied from 6.7 to 100 spikes/s. A gradual increase followed by a saturation of the EPSC amplitude can be observed with increasing *f*. The slopes of the I-O relation curves also increase with $V_m$. For example, the slopes at *f*=40 spikes/s are 1.55, 2.24 and 2.94 nA·s/spikes when $V_m$=–0.1, 0 and 0.1 V, respectively. Such results indicate that algebraic transformation on the neural I-O relationships is likely to be multiplicative operation. Furthermore, the I-O relation curves can be fitted by the empirical function written as:

$$R(d,m) = R(f,V_m) = \frac{A \cdot (1 + D \cdot V_m)}{1 + B \cdot \exp(-C \cdot f)},$$

where A, B, C and D are estimated to 129 nA, 2.46, 0.055 s/spikes and 3.1 $V^{-1}$, respectively. The experimental data was presented in **S4** of supplementary information. Therefore, the relationship can be written as a function of *f* multiplied by a function of $V_m$. Such results indicate the multiplicative operation was realized based on the rate coding scheme. Similarly, the neuronal arithmetic based on temporal-correlated coding scheme was also illustrated in **S5** in supplementary information. Systematic changes in the slope or the gain of the neural input-output relationship induced by multiplicative operation underlies a wide range of neural processes, including translation-invariant object recognition, visually guided reaching, collision avoidance, etc[46-48]. As the multiplicative operation can be realized by tuning the modulatory input, thus our artificial neuron could be potentially applied in pattern recognition and sensory processing applications.

In conclusion, proton conducting GO films were proposed as the electrostatic coupling electrolytes for multi-gate oxide-based neuron transistor fabrication. Paired-pulse facilitation, dendritic integration and orientation tuning functions were successfully emulated. Most importantly, neuronal gain control (arithmetic) in the scheme rate coding was also experimentally demonstrated. Since these functions are highly correlated to neural computations such as pattern recognition, sensory processing, etc., such neuron transistors provide the new-concept building blocks for neuromorphic cognitive systems. At present, the device size of the proof-of-principle artificial neurons is relatively large, and scaling the size down to 100 nm should be possible when an advanced photolithography process is used in the future. At the same time, flexible 3D integration should also be possible because all processes involved in the neuron transistor fabrication are performed at low temperature.

**Experiment**

*Preparation and Characterizations of Graphene Oxide (GO) films:* Firstly, 1.5 g

natural graphite flakes and 1.8 g $KNO_3$ was added into 69 mL of concentrated $H_2SO_4$ (98%) under stirring. After a few minutes, 12 g $KMnO_4$ was added slowly. The mixture was then heated to 40 °C and stirred for 6 hours. Subsequently, 120 mL water was added under vigorous stirring, resulting in a quick rise of the temperature to ~80 °C. The slurry was further stirred at this temperature for another 30 min. Afterwards, 300 mL water and 12 mL $H_2O_2$ solution (30 wt.%) were added in sequence to dissolve insoluble manganese species. The resulting graphite oxide suspension was washed repeatedly by plenty of water until the solution pH reached a constant value of ~5.0. Complete delamitation of graphite oxide into GO was achieved by ultrasonic treatment. Finally, the brown, homogeneous colloidal suspension of GO was obtained. X-ray photoelectron spectroscopy (XPS) measurements of GO films were carried out on AXIS UTLTRA DLD. For XPS measurement, GO solution were also spin-coated on a polished Si (100) wafer. The thickness of the GO films is measured by Stylus Profiler (Dektak150, Veeco).

*Fabrication of Neuron Transistors:* Firstly, GO suspensions (~6 mg/mL) were spin-coated onto ITO glass substrates and dried at 50 °C for 2 h. Then, a 30 nm thick patterned indium-zinc-oxide (IZO) channel layer was deposited on the GO films by radio-frequency (RF) magnetron sputtering with the aid of a nickel shadow mask. The sputtering was performed using an IZO ceramic target with a RF power of 100 W and a working pressure of 0.5 Pa. The channel width and length were 240 μm and 80 μm, respectively. Finally, patterned 100-nm thick Au source/drain and gate electrodes (240

μm×200 μm) were deposited through another nickel shadow mask by thermal evaporation method.

***Device Electrical Characterizations***: The frequency dependent capacitances of GO films were characterized by a Solartron 1260A Impedance/Gain-Phase Analyzer. The electrical measurements of the neuron transistors were performed on a semiconductor parameter characterization system (Keithley 4200 SCS) at room temperature with a relative humidity (RH) of 50 %.

**Acknowledgements**

This work was supported in part by the National Science Foundation for Distinguished Young Scholars of China (Grant No. 61425020), in part by the National Natural Science Foundation of China (11174300, 11474293), and in part by the Zhejiang Provincial Natural Science Fund (LR13F040001).

**Author contributions**

Q. W. and Y. S. conceived and designed the experiments; C. W., L. Z, Y. L. and P. F. fabricated the devices and performed electrical measurements. H. C., P. X., and Z. L prepared and characterized the GO films. The manuscript was written by C. W., L. Z., and Q. W.

**Competing financial interests**

The authors declare no competing financial interests.

**Figure legends**

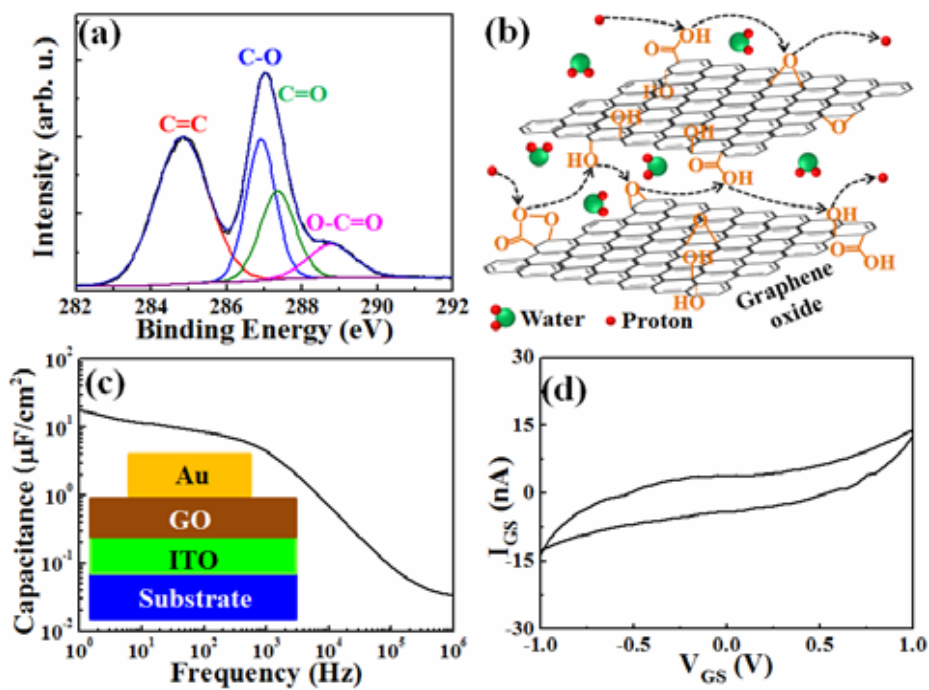

**Figure 1 | Characterization of the GO films. a,** XPS spectra of the GO film. b, Possible proton hopping paths in GO film. **c,** Frequency-dependent specific capacitance of the GO film. Inset: the Au/GO/ITO sandwich structure for measurement. **d,** The leakage current of the GO film in the voltage range from -1.0 to 1.0 V.

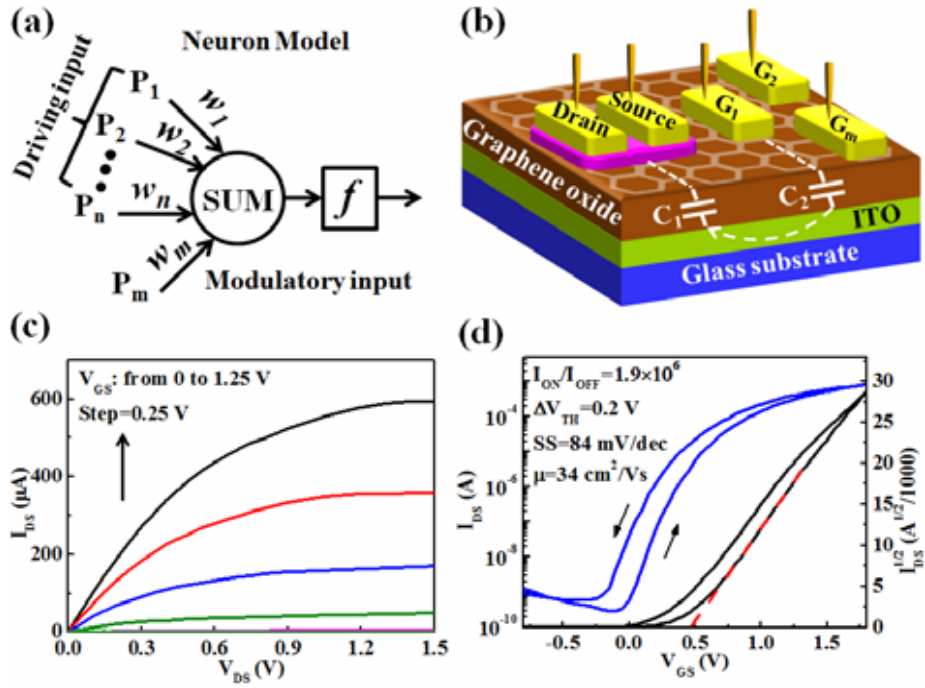

**Figure 2 | Structure and electrical performance of IZO-based neuron transistors gated by GO film.** **a,** A schematic image of an artificial neuron. The $w_i$ (i=1,2…n and m) denotes the synaptic weight of each input. **b,** The schematic diagram of an IZO-based neuron transistor gated by the GO electrolyte film on ITO glass substrate. **c,** The Output characteristics ($I_{DS}$ vs $V_{DS}$) of the neuron transistor. **d,** The transfer characteristics of the neuron transistor measured with $V_{DS}$=1.5 V.

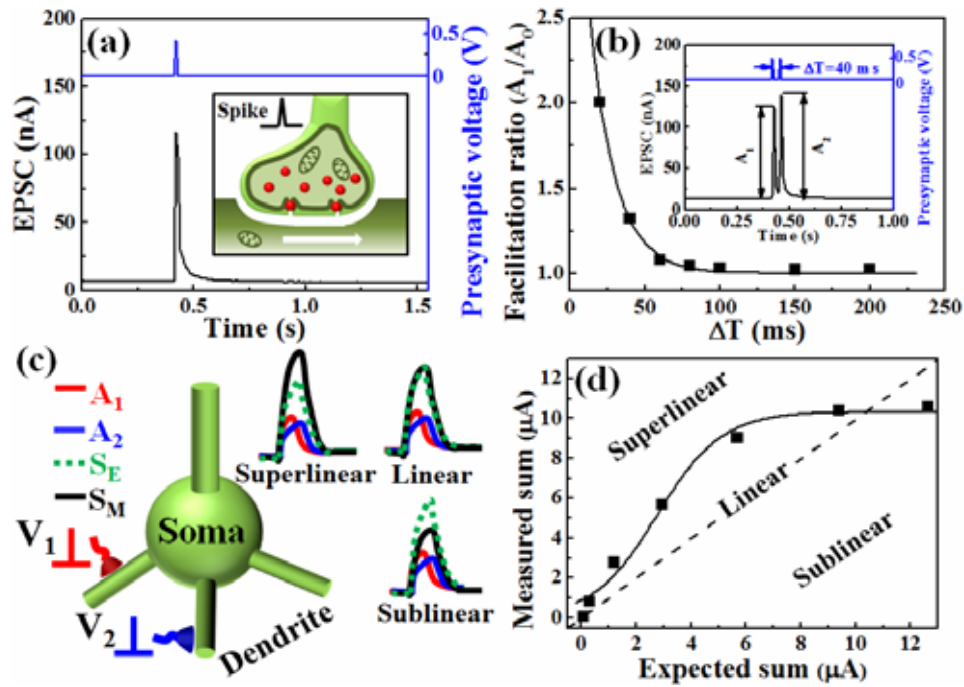

**Figure 3 | Synaptic plasticity and dendritic integration functions. a,** an EPSC stimulated by a presynaptic spike (0.5 V, 10 ms). The EPSC is measured with $V_{DS}$=0.1 V. Inset: a schematic diagram of the EPSC process in a biological synapse. **b,** The facilitation ratio ($A_2/A_1$) plotted as a function of time interval ($\Delta T$). Solid line: fitting curve as a guide to the eye. Inset: a typical EPSC stimulated by two successive presynaptic spikes with a time interval of 40 ms. **c,** The schematic diagram of the spatial summation with two spatial isolated synapses. The two simultaneous presynaptic spikes ($V_1$ and $V_2$) are applied on the two synapses, respectively. $A_1$ and $A_2$ are the EPSC amplitudes stimulated by individual spike. $S_E$ and $S_M$ are the arithmetic sum of the two amplitudes and measured sum stimulated by two simultaneously triggered spikes, respectively. **d,** The measured sum ($S_M$) plotted as a function of expected sum ($S_E$). Solid line: fitting curve as a guide to the eye.

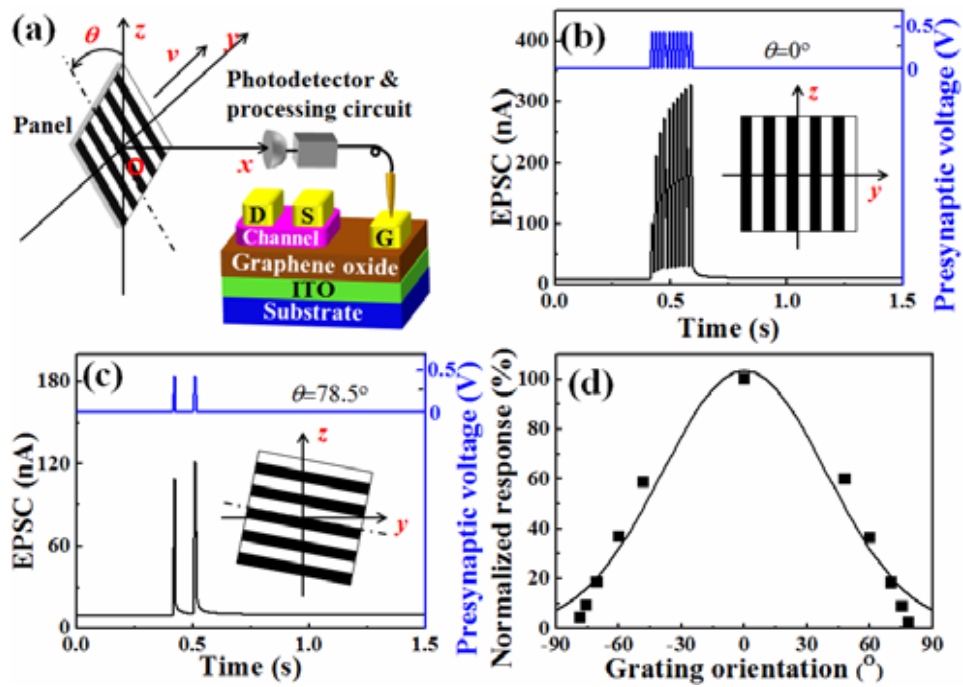

**Figure 4 | Orientation tuning experiments. a,** A schematic diagram showing the measurements for orientation tuning experiment. **b, c,** The EPSC recorded from a GO gated IZO-based neuron transistor when the orientation angles ($\theta$) of the panel are **(b)** $0°$ and **(c)** $78.5°$, respectively. **d,** The normalized visual response plotted as function of orientation angle. Solid line: fitting curve as a guide to the eye.

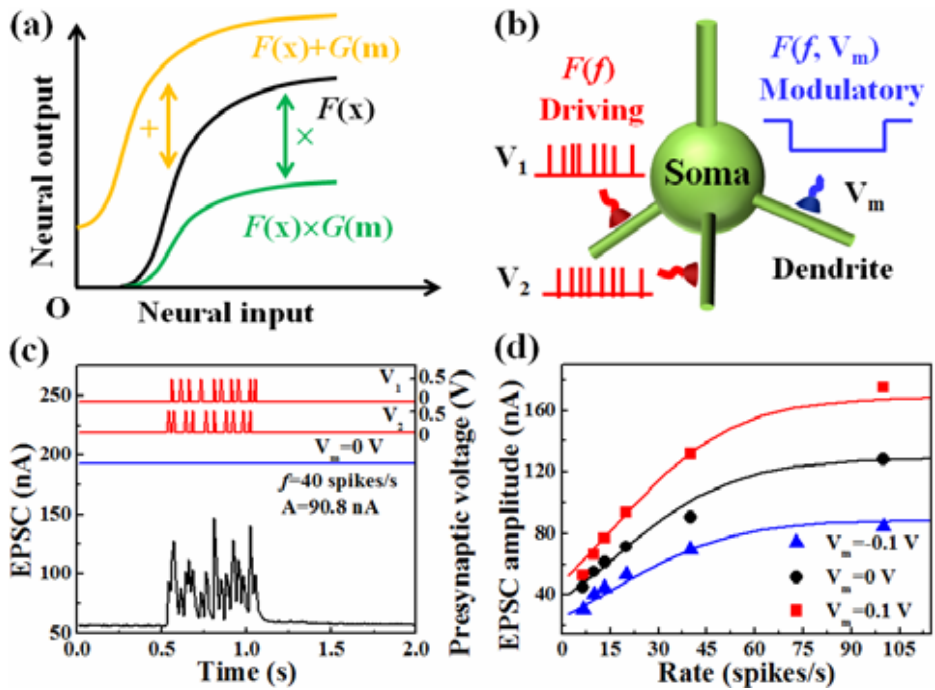

**Figure 5 | Neural arithmetic in the scheme of rate coding. a,** The schematic diagram of neuronal arithmetic when the modulatory input is independent of the neural input-output relationship of the driving input. **b,** The schematic diagram of rate coding scheme in a neuron with two driving synapses and a modulatory synapse. **c,** Neural output in response to two driving inputs and one modulatory input. The Poisson distributed driving inputs (0.5 V, 10 ms) triggered successively. The modulatory input is 0 V. **d,** The neuronal input-output relationships based on rate coding scheme were tuned by modulatory inputs of -0.1, 0 and 0.1 V, respectively.

Supplementary information

# Proton Conducting Graphene Oxide Coupled Neuron Transistors for Brain-Inspired Cognitive Systems

*Changjin Wan [1,2], Liqiang Zhu [2], Yanghui Liu [2], Ping Feng [1], Zhaoping Liu [2], Hailiang Cao [2], Peng Xiao [2], Yi Shi [1*], and Qing Wan,[1,2,*]*.

[1] School of Electronic Science & Engineering, and Collaborative Innovation Center of Advanced Microstructures, Nanjing University, Nanjing 210093, China

[2] Ningbo Institute of Material Technology and Engineering, Chinese Academy of Sciences, Ningbo 315201, China

[*] E-mail: wanqing@nju.edu.cn; yshi@nju.edu.cn

## S1. The experimental details of orientation tuning emulation.

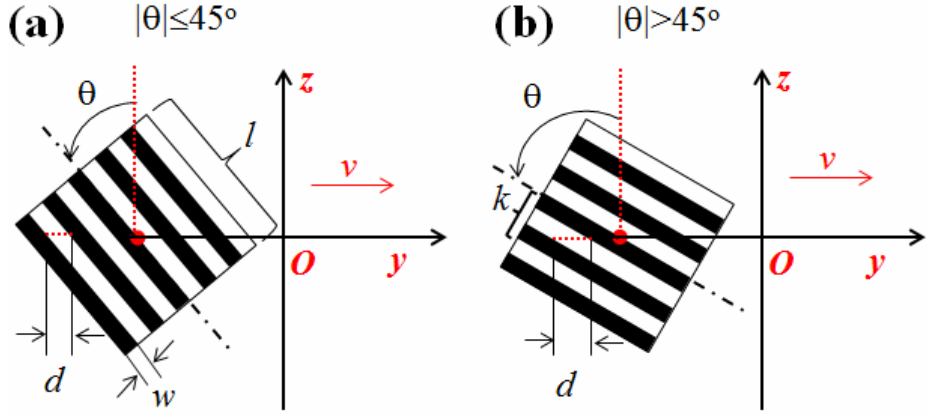

**Figure S1|** The figure schematically shows that the square panel with five pairs of black-white grating patterns moves along y axis in yoz plane (the background is white). The absolute values of the orientation angle are: (a) $|\theta| \leq 45°$ and (b) $|\theta| > 45°$, respectively. The width and length of each pattern (black or white) are $w$ and $l$ ($w$: $l$=1:10), respectively. The coordinate of the photodetector is (x, 0, 0) and that of the panel center is (0, y, 0). The velocity of the panel in y axis is $v$. $\theta$ could be altered between -90° and +90°. For each orientation, the panel is moved from side to side. Each time when the edge of the grating pattern move across the coordinate origin is detected by the photodetector, the processing circuit would provide a voltage pulse (0.5 V, 10 ms) to the presynaptic terminal (gate) of the GO gated IZO neuron transistor.

If $|\theta| \leq 45°$, ten spikes will be triggered when the square panel moved from side to side. Thus the frequency of the presynaptic spikes is: $f = v/d = v \cdot \cos\theta/w$.

If $|\theta| > 45°$, the spike numbers is dependent on the value of $2 \cdot k/w$, where $k = |5w/\tan\theta|$. In this work, the spike numbers (N) are determined by the followed equation:

$$N = \begin{cases} [|10/\tan\theta|], & |10/\tan\theta| - [|10/\tan\theta|] \leq 0.5 \\ [|10/\tan\theta|]+1, & |10/\tan\theta| - [|10/\tan\theta|] > 0.5 \end{cases},$$

where the [m] means the integer part of m.

Similarly, the frequency of the presynaptic spikes is: $f = v \cdot \cos\theta/w$.

In the simulation experiments, the orientation angles are 0°, ±48.2°, ±60°, ±70.5°, ±75.5° and ±78.5°, respectively. The value of $v/w$ is set to 20 ms in this work. In this case, the frequencies of the presynaptic spikes are 50, 33.3, 25, 16.7, 12.5 and 10 Hz, respectively. And the numbers of the presynaptic spikes are 10, 9, 6, 4, 3 and 2, respectively.

**S2. Supplementary animations of orientation tuning experiments.**

The GIF format files titled "angle-0.gif", "angle-48-2.gif", "angle-60-0.gif", "angle-70-5.gif", "angle-75-5.gif" and "angle-78-5.gif" illustrate the dynamic processes when the square panel move along the y axis with different orientations angle of 0°, 48.2°, 60°, 70.5°, 75.5° and 78.5°, respectively. The spikes generated at each time when the edge of the grating pattern was detected were also shown in the figures.

**S3. The protocol of Poisson-distributed presynaptic spike train.**

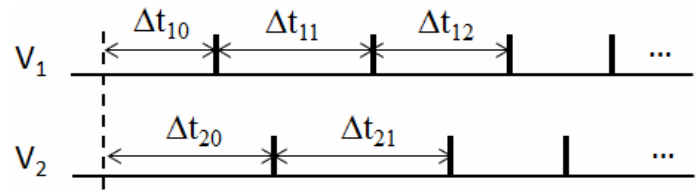

**Figure S2** | The schematic diagram illustrated the protocol of Poisson distributed spike trains. The dashed line is the virtual time origin. The two horizontal lines is time axis and each short vertical line represents a presynaptic spike (0.5 V, 10 ms).

The two presynaptic spike trains were applied on $G_1$ and $G_2$, respectively. Each train consists of ten presynaptic spikes. For each train, the time interval ($\Delta t_{ij}$) between spikes is Poisson distributed:

$$P(x = \Delta t_{ij}) = \frac{\lambda^k}{k!} e^{-\Delta t_{ij}}, \quad i = 1 \quad or \quad 2, \quad and \quad j = 0, 1, ..., 9,$$

where $\lambda$ and $k$ are the expectation (or variance) and number of the time intervals,

respectively. The first time interval ($\Delta t_{10}$ and $\Delta t_{20}$) of each train is the time interval between the first spike and the virtual time origin as shown in Fig. S2. In our case, $\lambda$=20, 50, 100, 150, 200 and 300 ms were used. As the spike number of each spike train is 10, thus k=10. According to the definition of Poisson distribution, the expectation of the random sequence is equal to $\lambda$. Therefore the rate of the two spike trains is estimated to 2000/$\lambda$ spikes/s in this work. The Poisson distributed random numbers were generated by matlab software. As the time resolution of the Keithlay 4200 used in this work is ~10 ms, the random numbers are set as multiples of ten milliseconds.

## S4. Experimental data for neural arithmetic based on rate coding scheme.

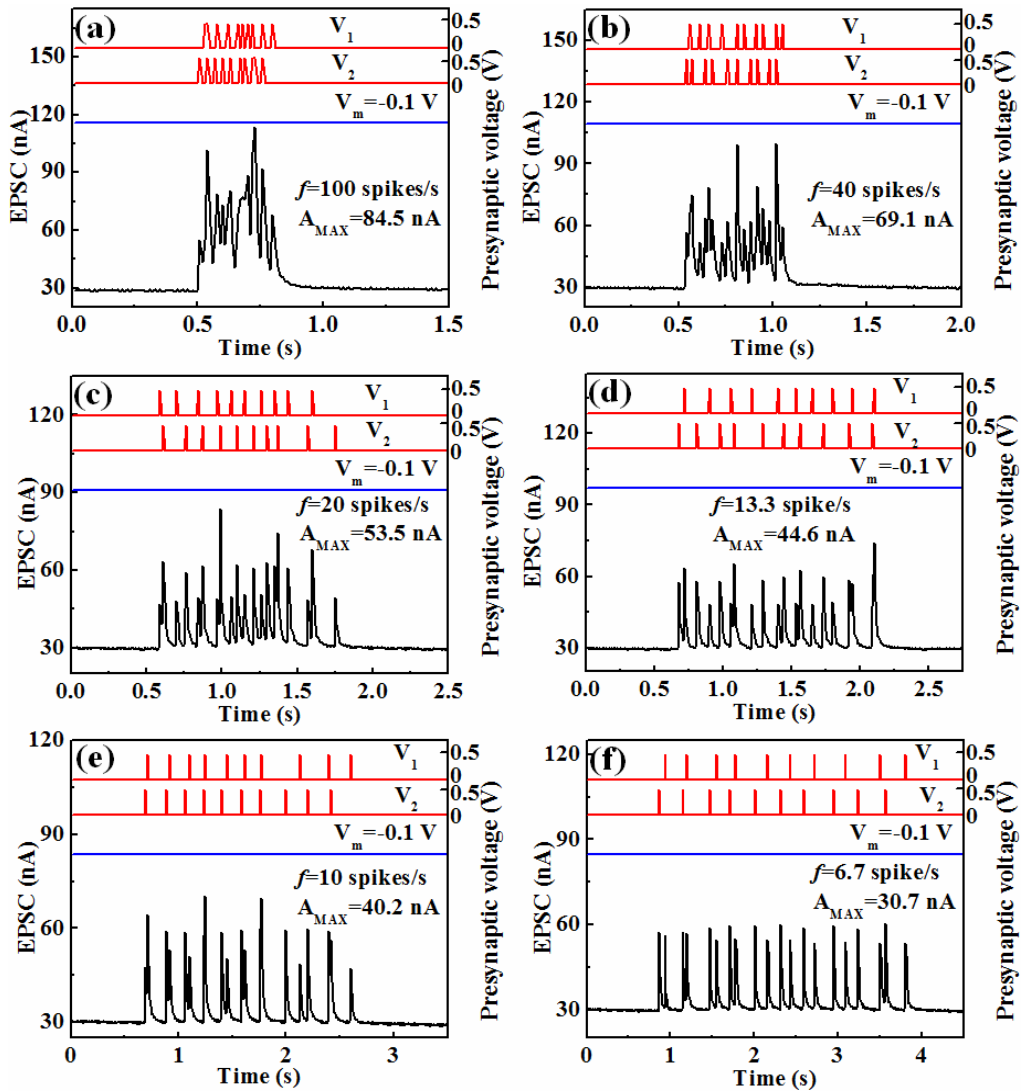

**Figure S3|** The EPSC recorded in response to the rate of the driving inputs were (a)

100, (b) 40, (c) 20, (d) 13.3, (e) 10 and (f) 6.7 spikes/s, respectively. The modulatory input voltage is -0.1 V.

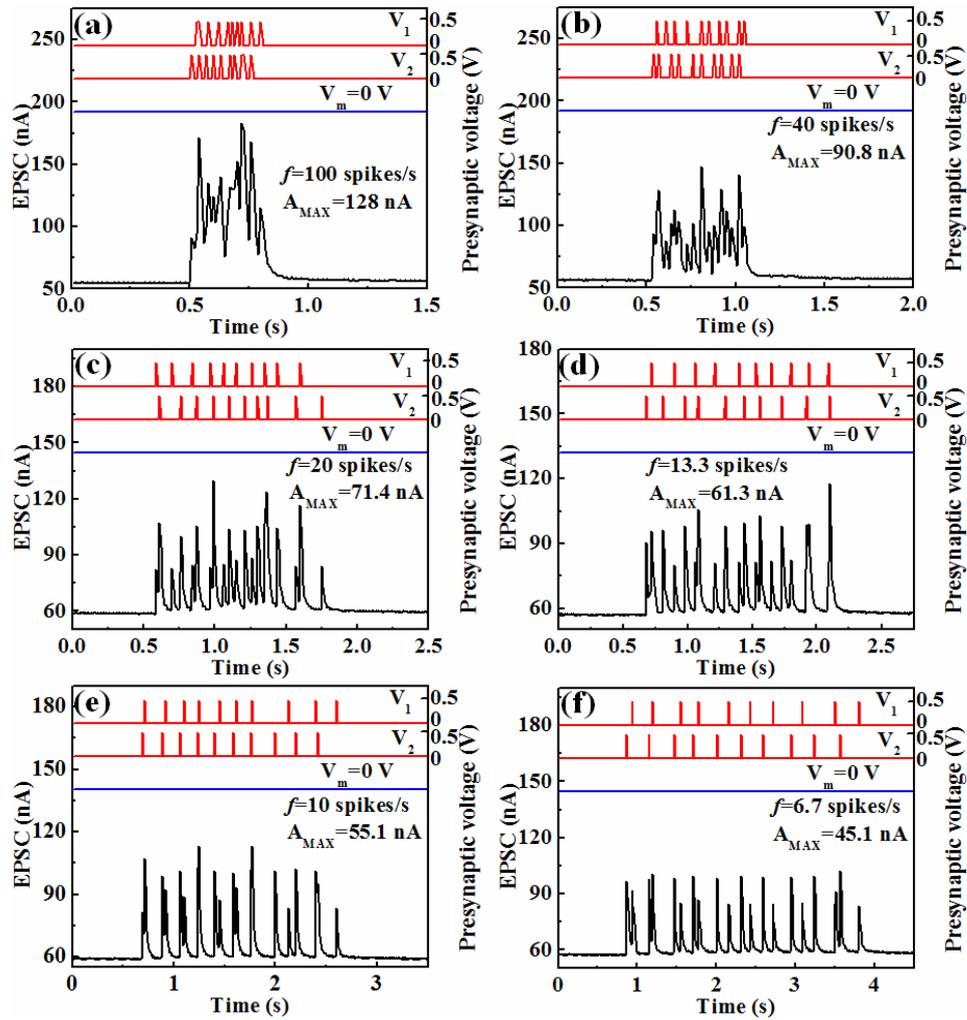

**Figure S4|** The EPSC recorded in response to the rate of the driving inputs were (a) 100, (b) 40, (c) 20, (d) 13.3, (e) 10 and (f) 6.7 spikes/s, respectively. The modulatory input voltage is 0 V.

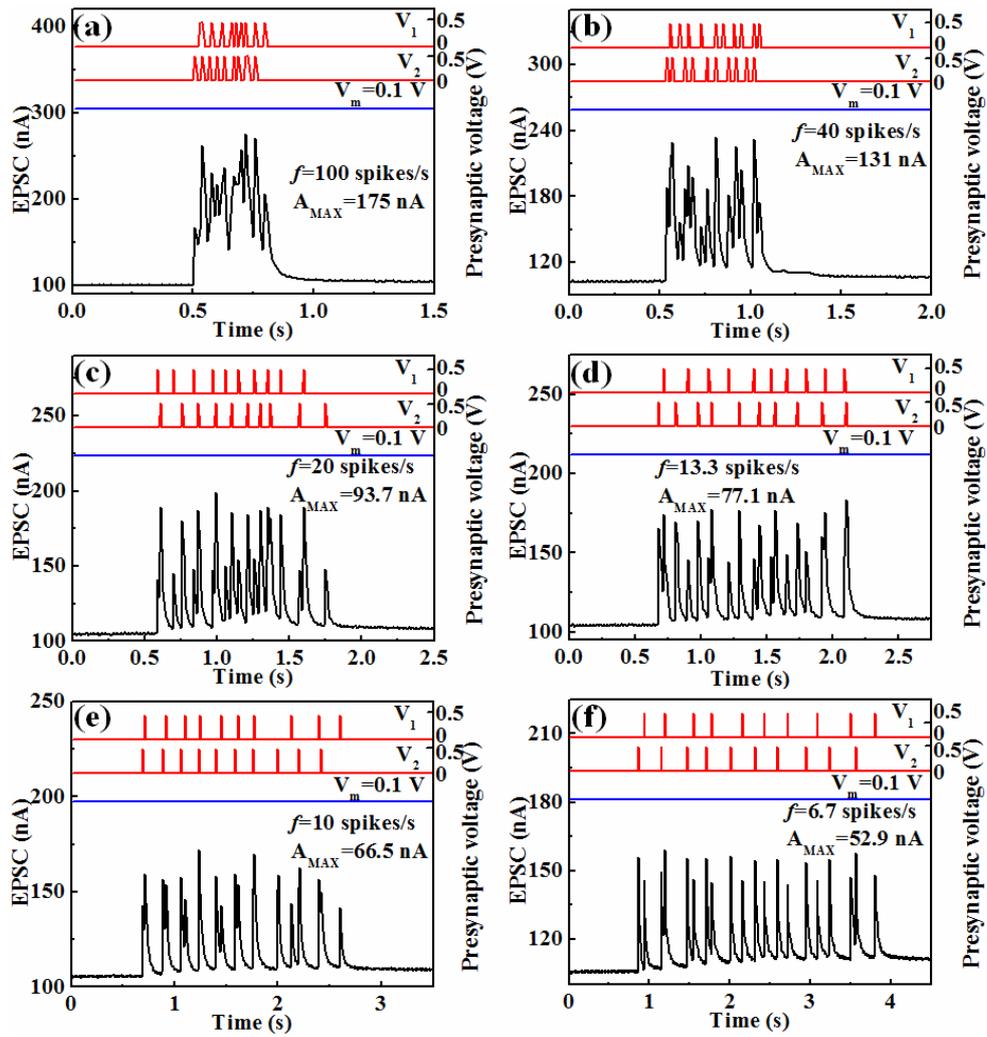

**Figure S5|** The EPSC recorded in response to the rate of the driving inputs were (a) 100, (b) 40, (c) 20, (d) 13.3, (e) 10 and (f) 6.7 spikes/s, respectively. The modulatory input voltage is 0.1 V.

## S5 Multiplicative operation based on temporal-correlated coding

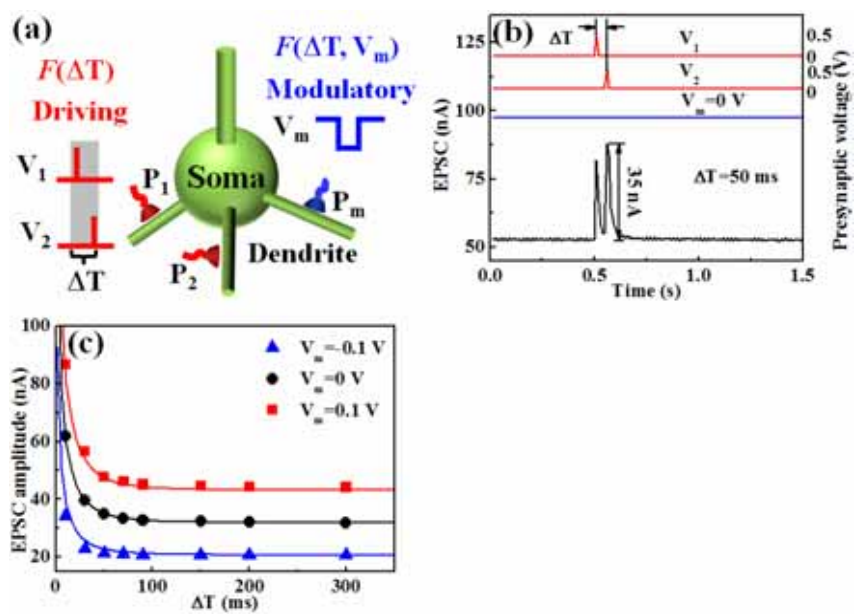

**Figure S6| a,** The schematic diagram of temporal-correlated coding. **b,** EPSCs recorded in response to two driving inputs and one modulatory input. The driving input (0.5 V, 10 ms) triggered successively with a time interval of 50 ms. The modulatory input is 0 V. **c,** The neuronal input-output relationships based on temporal-correlated coding were tuned by different modulatory inputs of -0.1, 0 and 0.1 V, respectively.

Figure S6a schematically shows the temporally correlated coding scheme in which information is encoded by the correlations in spike timing. In our case, driving input parameter is defined as the time interval between the two presynaptic inputs and the modulatory input is defined as the modulatory voltage, i. e. d=ΔT and m=$V_m$. Figure S6b shows a typical neural output in response to two driving inputs (0.5 V, 10 ms) with time interval of 50 ms and a modulatory input of 0 V. The amplitude is estimated to 35 nA. Figure S6c systematically shows the neural I-O relation curves under modulatory input voltages of -0.1, 0 and 0.1 V, respectively. For each curve, the time intervals are varied from 10 ms to 300 ms. The neural output decreased gradually with increasing ΔT. For example, when d=10 ms and m=0.1 V, the neural output is ~87 nA, and when d=200 ms and m=0.1 V, the neural output is ~44 nA. The observed results can be understood by considering the migration of ions in the GO film and the electrostatic coupling effect of the EDL. A preceded positive presynaptic spike accumulates ions around the IZO channel. After the spike, protons will migrate back to their equilibrium positions gradually, which enhance the EPSC amplitude triggered by the following spike.

The slopes of the neural I-O relation curves also increase with $V_m$; for example, the slopes at $\Delta T=50$ ms are −79, −121 and −167 pA/ms when $V_m=$−0.1, 0 and 0.1 V, respectively. Such results indicate that the most likely algebraic transformation on the neural I-O relationships is the multiplicative operation. Furthermore, the I-O relation curves can be fitted by the empirical function written as:

$$R(d,m) = R(\Delta T, V_m) = \left\{ A \cdot exp\left[-\left(\frac{\Delta T}{B}\right)^C\right] + D \right\} \cdot (1 + E \cdot V_m),$$

where A, B, C, D and E are estimated to 216 nA, 2.34 ms, 0.47, 32 nA and 3.5 $V^{-1}$, respectively. Therefore, the relationship can be written as a function of $\Delta T$ multiplied by a function of $V_m$. Such results indicate the multiplicative operation was realized based on the temporal-correlated coding scheme.